\documentstyle[12pt]{article}
\makeatletter
\@addtoreset{equation}{section}
\makeatother

\def\ftoday{{\sl {Le \number\day \space\ifcase\month 
\or janvier\or f\'evrier\or mars\or avril\or mai
\or juin\or juillet\or ao\^ut\or septembre\or octobre
\or novembre \or d\'ecembre\fi\space \number\year}}}    
\def\ptoday{{\sl {\number\day \space de\space \ifcase\month 
\or janeiro\or fevereiro\or mar{\c c}o\or abril\or maio
\or junho\or julho\or agosto\or setembro\or outubro
\or novembro \or dezembro\fi\space de\space \number\year}}}    
\def\gtoday{{\sl {Den \number\day. \ifcase\month 
\or Januar\or Februar\or M\"arz\or April\or Mai
\or Juni\or Juli\or August\or September\or Oktober
\or November \or Dezember\fi\space \number\year}}}    
\def\today{{\sl {\ifcase\month
\or January\or February\or March\or April\or May
\or June\or July\or August\or September\or October
\or November \or December\fi \space\number\day,\space 
                                            \number\year}}}
\newcommand{\journal}[4]{{\em #1~}#2\,(#3)\,#4}

\newcommand{\hpa}{\journal {Helv. Phys. Acta}}

\newcommand{\ijmp}{\journal {Int. J. Mod. Phys.}}

\newcommand{\pr}{\journal {Phys. Rev.}}

\newcommand{\jmp}{\journal {J. Math. Phys.}}

\newcommand{\cmp}{\journal {Commun. Math. Phys.}}

\newcommand{\cqg}{\journal {Class. Quantum Grav.}}

\newcommand{\np}{\journal {Nucl. Phys.}}
\newcommand{\pl}{\journal {Phys. Lett.}}

\newcommand{\nc}{\journal {Nuovo Cimento}}

\renewcommand{\d}{\delta}         
\newcommand{\e}{\varepsilon}

\newcommand{\la}{\lambda}        
\newcommand{\m}{\mu}
\newcommand{\n}{\nu}
\newcommand{\om}{\omega}

           \newcommand{\F}{{\Phi}}
\newcommand{\vf}{{\varphi}}

\renewcommand{\AA}{{\cal A}}

\newcommand{\LL}{{\cal L}}

\newcommand{\es}{\\[3mm]}

\newcommand{\sla}{\raise.15ex\hbox{$/$}\kern -.57em} 
\newcommand{\Sla}{\raise.15ex\hbox{$/$}\kern -.70em}

\newcommand{\lp}{\left(}\newcommand{\rp}{\right)}

\newcommand{\complex}{{\kern .1em {\raise .47ex
\hbox {$\scriptscriptstyle |$}}
    \kern -.4em {\rm C}}}
\newcommand{\real}{{{\rm I} \kern -.19em {\rm R}}}
\newcommand{\rational}{{\kern .1em {\raise .47ex
\hbox{$\scripscriptstyle |$}}
    \kern -.35em {\rm Q}}}
\renewcommand{\natural}{{\vrule height 1.6ex width
.05em depth 0ex \kern -.35em {\rm N}}}

\newcommand{\half}{\dfrac{1}{2}}
\newcommand{\pa}{\partial}

\newcommand{\dfud}[2]{{\displaystyle{\frac{\delta #1}{\delta #2}}}}
\newcommand{\dfrac}[2]{{\displaystyle{\frac{#1}{#2}}}}
\newcommand{\dsum}[2]{\displaystyle{\sum_{#1}^{#2}}}

\newcommand{\twiddle}{\lower.9ex\rlap{$\kern -.1em\scriptstyle\sim$}}

\newcommand{\equ}[1]{(\ref{#1})}
\newcommand{\eq}{\begin{equation}}
\newcommand{\eqn}[1]{\label{#1}\end{equation}}
\newcommand{\eea}{\end{eqnarray}}
\newcommand{\eqa}{\begin{eqnarray}}
\newcommand{\eqan}[1]{\label{#1}\end{eqnarray}}
\newcommand{\ba}{\begin{array}}
\newcommand{\ea}{\end{array}}
\newcommand{\eqac}{\begin{equation}\begin{array}{rcl}}
\newcommand{\eqacn}[1]{\end{array}\label{#1}\end{equation}}

\newcommand{\nablag}{\nabla^{({\rm g})}}
\newcommand{\dse}{\d^{\,{\rm S}}_{(\e)}}
\newcommand{\sqg}{\sqrt{-g}}
\newcommand{\BC}{{\bar C}}
\newcommand{\bxi}{{\bar\xi}}
\newcommand{\bom}{{\bar\om}}
\newcommand{\dintm}{\displaystyle{\int_M}}
\begin{document}

{\hfill UFES-DF-OP2000/1}

{\hfill hep-th/0005011}

\begin{center}
{{\LARGE {\bf Ghost Equations and\es
 Diffeomorphism Invariant Theories}}}

\vspace{7mm}

{\large 
Olivier Piguet\footnote{{{Supported in part by the 
{\it Conselho Nacional de
Desenvolvimento Cient\'\i fico e Tecnol\'ogico (CNPq -- Brazil)}.}}}}
\vspace{1mm}

 {\it Universidade Federal do Esp\'{\i}rito Santo (UFES)
, \\CCE, Departamento de F\'{\i}sica,\\ Campus Universit\'ario
de Goiabeiras - BR-29060-900 - Vit\'oria - ES - Brasil.}

{\tt E-mail: piguet@cce.ufes.br}

May, 2000 (revised\footnote{Sign in eq. \equ{susy-cond} corrected;
references added.} July, 2000)
\end{center}

{\small 
\noindent
{\bf Abstract} Four-dimensional Einstein gravity in the Palatini first order formalism is
shown to possess a vector supersymmetry of the same type as found in
the topological theories for Yang-Mills fields. A peculiar feature of
the gravitationel theory, characterized by diffeomorphism invariance, is a 
direct link of vector supersymmetry with the field equation of motion
for the Faddeev-Popov ghost of diffeomorphisms. }

\section{Introduction}



In the first order formalism of four-dimensional 
gravitation theory~\cite{peldan}, 
the independent dynamical variables 
are the vierbein 1-form $E$ (giving the metric $G$) 
and the connection 1-form $A$. 
The vierbein
dependence of the connection is given by the field equation with respect
to $A$, whereas Einstein equation results from the field equation with
respect to $E$. The action will be written, 
following~\cite{baez-knots,baez-foam}, 
in a ``topological'' form, i.e. in such a way that
it can be interpreted as an action of the 1-form fields $E$ and $A$ on a
differentiable manifold $M$, without reference to any a-priori background
metric. The latter point is known~\cite{witten} to be an essential 
characteristic of  
topological theories, and trying to exploit this feature belongs to
 the spirit of the modern attempts 
towards a construction of quantum gravity 
(see~\cite{gaul-rovelli,rovelli-living} for 
reviews and further references). 

Since the theory possesses two local symmetries 
-- the diffeomorphism and local
Lorentz invariances -- one has to perform a gauge fixing for both.
We will choose a gauge fixing of the Landau type, within
the BRST formalism~\cite{brst,alg-ren}. Much in the same way 
as in topological theories, this requires the introduction of a
nondynamical, background metric $g$. This construction closely
parallels the one performed for the Chern-Simons theory 
in~\cite{cs-lucch-pig}. It should be clear that the
background metric, being introduced only in the gauge fixing part of the
theory, should not affect in any way the physical 
outcome, as it has been shown for instance in~\cite{cs-lucch-pig} 
for the  -- perturbative -- quantum version of the
Chern-Simons theory.

An interesting features of topological theories such as 
Chern-Simons or $BF$ theory, is the presence of a
``vector supersymmetry'' -- a supersymmetry whose generator is a vector
valued operator~\cite{vec-susy}. 
In case the manifold admits isometries generated by 
Killing vectors -- e.g. the space-time translations 
if the background metric is 
flat -- the vector supersymmetry is a symmetry of the gauge-fixed
action. It happens that its generator together with the
BRST symmetry generator form an algebra which closes on the 
generators of the isometries of the translation type~\cite{cs-lucch-pig}. 
The vector supersymmetry
has been shown to play a key role in the ultraviolet finiteness of the
topological 
theories~\cite{UV-finiteness-vector-SUSY,cs-lucch-pig}.

Another interesting features -- actually shared by any gauge theory,
provided its
gauge fixing be of the Landau type -- is the so-called 
ghost equation~\cite{ghost-eq,alg-ren}, 
which restricts the coupling of the
ghosts and implies the nonrenormalization of their
field amplitude\footnote{A review 
of the properties of topological theories mentionned above may
be found in Chapters 6 and 7 of the book~\cite{alg-ren}.}.

The purpose of the present note is to show the existence of such a
vector supersymmetry for Einstein gravity in the Palatini formalism. We
shall in fact see that the vector supersymmetry 
is a direct consequence of
the field equation of the Feynman-Dewitt-Faddeev-Popov 
ghost~\cite{ghost} associated to
diffeomorphism invariance. On the other hand, the ghost 
equation related
to local Lorentz invariance will be seen to be algebraically 
associated with rigid Lorentz invariance. 
We shall also see that this
supersymmetry, like in topological theories, yields the Sorella 
operator~\cite{sorella-delta} $\d$ used in order to solve the 
BRST cohomology and to construct the
invariants of the theory. The operator $\d$ has been given for gravity 
in~\cite{brs-grav-wien1,brs-grav-wien2}.

To the contrary of the topological theories of a Yang-Mills
connection (Chern-Simons or BF), 
where the supersymmetry generators are 
the components of a vector and where the superalgebra closes 
on the translations, Einstein gravity in the Palatini 
formalism studied in the present paper will be seen to
admit a supersymmetry
possessing generators which are components of 
one vector and one antisymmetric 
tensor, the full algebra 
containing now all the ten
Poincar\'e generators -- in the 
case of a flat background metric at least\footnote{Such a tensor 
supersymmetry has been pointed out in~\cite{emery}, in the case of a 
Chern Simons model in a gravitational background in the vielbein 
formalism.}. 

Although the present work will only be concerned with the 
classical aspects of the theory, the results are of interest
since, as we already said, they reveal the link between the construction 
of the observables via the $\delta$ operator of 
Sorella~\cite{sorella-delta,brs-grav-wien1,brs-grav-wien2}, 
on one hand, and
the gauge fixing, through the ghost equation, on the other hand.

\section{Symmetries, Gauge Fixing and BRST Invariance}

The Einstein gravity Lagrangian in the first order formalism of 
Palatini~\cite{peldan} may be written as~\cite{baez-knots,baez-foam}:
\eq
S_{\rm inv} =
 \dfrac{1}{4}\dintm 
\e_{IJKL} E^I\wedge E^J \wedge F^{KL}(A) + S_{\rm matter}(E,A,\F)\ .
\eqn{inv-action}
The integral is taken over some differentiable 4-manifold $M$, 
$E^I$ is the vierbein 1-form, with $I=0,\cdots,3$ a tangent 
plane Lorentz index. $F^{KL}$ is the curvature 2-form
\eq
F^{IJ}(A) = d A^{IJ} + A^{IK}\wedge A_K{}^J
\eqn{curv} 
of the Lorentz
connection\footnote{If the connection is self-dual, \equ{inv-action}
is the Ashtekar action~\cite{ashtekar,baez-knots}.}
$A^{IJ}$, the latter being taken as an independant  
variable\footnote{In a particular coordinate frame with 
$x =$ $(x^\m,\,\m=0,\cdots,3)$, $E^I = E_\m^I dx^\m$,
$F^{IJ} = \half F_{\m\n}^{IJ} dx^\m\wedge dx^\n$, etc.}. 
$\e_{IJKL}$ is the rank four totally antisymmetric tensor, normalized by
$\e_{0123}=1$. In the following, the  exterior multiplication
symbol $\wedge$ will be omitted. 
$S_{\rm matter}$ is some action for minimally coupled matter fields $\F$,
which we don't need to specify. We shall in fact omit this 
part in the following, for the sake of simplicity.

The field equations given by the variations of this action read
\eq\ba{l}
\dfud{S_{\rm inv}}{E^I} = \half\e_{IJKL} E^J F^{KL} \ , \es
\dfud{S_{\rm inv}}{A^{IJ}} = \e_{IJKL} E^K DE^L \ ,
\ea\eqn{eq-motion}
where $D$ is the covariant exterior derivative:
$DE^I = dE^I - A^I{}_J E^J$.
It is known~\cite{peldan,baez-foam} 
that they lead to the usual specification of a torsion-free connection
function of the vierbein\footnote{This is true for pure gravity. 
In case of coupling with matter, the second field equation does not 
automatically lead to a vanishing torsion~\cite{peldan}. One may then choose
to stay with a non-vanishing action, or to impose a supplementary
condition.}
and to the Einstein equation, in a Riemanian
space-time with metric
\eq
G_{\m\n}= E_\m^I E_\n^J \eta_{IJ}\ ,
\eqn{metric}
where $\eta_{IJ}$ is the Minkowsky 
metric\footnote{We consider a Lorentzian signature. But everything 
applies as well to the Euclidean case.} ${\rm diag}(1,-1,-1,-1,)$, used to 
lower and rise the tangent space indices $I,J,\cdots$.
 
The action \equ{inv-action} is invariant under the diffeomorphisms,
written in infinitesimal form, the infinitesimal parameter being a 
vector field $\xi$:
\eq
\d_{(\xi)} \vf = \LL_\xi \vf\ ,\quad \vf = E^I,\,A^{IJ}\ ,
\eqn{diff}
where $\LL_\xi$ is the Lie derivative along the vector $\xi$. 
It is also invariant under 
the local Lorentz transformations -- 
written in infinitesimal form,
with local parameters $\om^{IJ}=-\om^{JI}$:
\eq\ba{l}
\d_{(\om)} E^I = \om^I{}_J E^J\ ,\es
\d_{(\om)} A^{IJ} =  d\om^{IJ} + \om^I{}_K A^{KJ} + \om^J{}_K A^{IK}\ .
\ea\eqn{loc-lor}
In view of the gauge fixing procedure it is convenient to  express these
local invariances in the form of a nilpotent BRST operation $s$ defined 
by~\cite{BRS-grav}:
\eq\ba{l}
s E^I = \LL_\xi E^I + \om^I{}_J E^J\ ,\es
s A^{IJ} = \LL_\xi A^{IJ}
  + d\om^{IJ} + \om^I{}_K A^{KJ} + \om^J{}_K A^{IK}\ ,\es
s \xi = \half \{\xi,\xi\} \ ,\quad  
(\, \mbox{or: }s\xi^\m = \xi^\la\pa_\la \xi^\m\,) \ ,\es
s \om^I{}_J = \LL_\xi \om^I{}_J + \om^I{}_K\om^K{}_J\ ,
\ea\eqn{BRS}
with $s^2=0$. The infinitesimal parameters $\xi^\m(x)$ - the components of
the vector $\xi$ -- and $\om^I{}_J(x)$ are now Grassmann  
(i.e. anticommuting) number fields -- the Faddeev-Popov ghosts.
The bracket $\{\ ,\ \}$ is the Lie 
bracket\footnote{In a particular coordinate frame, the Lie bracket of 2
vectors $u$, $v$ takes the form
\[
\{u,v\}^\m = u^\la\pa_\la v^\m \pm v^\la\pa_\la u^\m\,,
\]
with the sign $+$ is both $u$ e $v$ are odd, and the sign $-$ otherwise.
 Even (odd) refers to the commuting (anticommuting) character of the 
object.}.

In order to gauge fix the theory with respect to its local symmetries --
diffeomorphism and local Lorentz invariances -- we introduce 
antighosts\footnote{Despite of an unfortunate but usual terminology,
the antighosts are independent of the ghosts~\cite{dewitt}.}
$\bxi_I$, $\bom_{IJ}$ and Lagrange multipliers $\la_I$, $b_{IJ}$, with
the following nilpotent BRST transformations:
\eq
s\bxi_I=\la_I\ ,\quad s\la_I=0\ ,\quad\quad
s\bom_{IJ}=b_{IJ}\ ,\quad sb_{IJ}=0\ .
\eqn{BRS-fix}
The gauge fixing part of the action is then defined as:
\eq\ba{l}
S_{gf} = -s\dintm d^4x \sqrt{-g} g^{\m\n}\lp \pa_\m\bxi_I E^I_\n + 
  \half \pa_\m\bom_{IJ} A^{IJ}_\n \rp \es
\phantom{S_{gf} } = -\dintm d^4x \sqrt{-g} g^{\m\n}\lp \pa_\m\la_I E^I_\n + 
  \half \pa_\m b_{IJ} A^{IJ}_\n \rp \es
\phantom{S_{gf}=} + \dintm d^4x \sqrt{-g} g^{\m\n}\lp \pa_\m\bxi_I sE^I_\n + 
  \half \pa_\m\bom_{IJ} sA^{IJ}_\n \rp\ ,
\ea\eqn{gf-action}
which is automatically BRST invariant. Note that in order to contract the
world indices $\m$, $\n$ we have introduced a (BRST-invariant)
background metric
$g_{\m\n}$ -- not to be confounded with the physical, 
dynamical metric $G_{\m\n}$
defined in \equ{metric}.

This particular gauge fixing, which is of the Landau type, is
completely determined by BRST invariance and by the ``gauge 
conditions'' -- i.e. by the field equations for the Lagrange 
multipliers:
\eq 
\dfud{S}{\la_I} = \pa_\m\lp \sqg g^{\m\n}E_\n^I \rp\ ,\quad
\dfud{S}{b_{IJ}} = \pa_\m\lp \sqg g^{\m\n}A_\n^{IJ} \rp \ ,
\eqn{gauge-cond}
where $S$ is the total action
\eq
S = S_{\rm inv} + S_{\rm gf}\ ,
\eqn{tot-action}
which is BRST invariant by construction: $\ sS=0$.

On the other hand, the field equations for the ghosts $\xi$ and 
$\om$ are
\eq\ba{l}
\dfud{S}{\xi^\m} =  \lp
-\sqg g^{\n\la} \pa_\la\bxi_I\pa_\m E_\n^I  +
\pa_\n \lp \sqg g^{\n\la} \pa_\la\bxi_I  E_\m^I \rp \rp \es
\phantom{\dfud{S}{\xi^\m} =  \lp\right.}
-\half\sqg g^{\n\la} \pa_\la\bom_{IJ}\pa_\m A^{IJ}_\n  +
\half\pa_\n \lp \sqg g^{\n\la} 
          \pa_\la\bom_{IJ}  A^{IJ}_\m\rp   \ ,\es
\dfud{S}{\om^{IJ}} = 
  -\sqg g^{\m\n} \lp \pa_\m \bxi_I E_{\n J} 
     + \pa_\m \bom_{IK} A_{\n J}{}^K
     -(I \leftrightarrow J) \rp 
 + \pa_\m \lp\sqg g^{\m\n} \pa_\n \bom_{IJ}\rp \ .
\ea\eqn{ghosteqs}

\section{Ghost Equation and Vector Supersymmetry}
Let us introduce the condensed notation
\eq\ba{ll}
\{\AA^i_\m,\,i =1,2\} = \{E_\m^I,\,A_\m^{IJ}\}\ ,\quad
     &\es
\{\BC_i,\,i =1,2\} = \{\bxi_I,\,\bom_{IJ}\}    \ ,\quad  
       &\{B_i,\,i =1,2\}= \{\la_I,\,b_{IJ}\}\ ,
\ea\eqn{notation}
under which the gauge conditions \equ{gauge-cond} and the 
equation for 
the ghost $\xi$ \equ{ghosteqs} now read\footnote{Summations 
$\sum_I$ and $\frac{1}{2}\sum_{IJ}$ over repeated indices are implicit.} 
\eq 
\dfud{S}{B_i} = \pa_\m\lp \sqg g^{\m\n}\AA^i_\n \rp\ ,
\eqn{g-cond}
\eq
\dfud{S}{\xi^\m} = \dsum{i=1,2}{} \lp
-\sqg g^{\n\la} \pa_\la\BC_i\pa_\m\AA^i_\n  +
\pa_\n \lp \sqg g^{\n\la} \pa_\la\BC_i \AA^i_\m\rp \rp \ .
\eqn{xi-eq}
What we claim here is that, under circonstances to be 
specified later on, the theory is invariant under the 
{\it vector supersymmetry transformations} (we use the 
notation \equ{notation})
\eq\ba{l}
\dse \xi^\m = \e^\m\ ,\es
\dse B_i = \e^\m\pa_\m \BC_i\ , \es
\dse \vf = 0\ , \quad \vf \not= \xi^\m\,,\ B^i  \ ,
\ea\eqn{vec-susy}
where the infinitesimal parameter 
$\e^\m$ is a vector field -- taken as commuting, to the 
contrary of $\xi^\m$, so that the supersymmetry operator $\dse$ is an
antiderivation. The latter together 
with the BRST operator $s$ obey the superalgebra anticommutation
relations
\eq
s^2\vf=0\ ,\quad (\dse)^2\vf=0\ ,\quad
\{s,\dse\}\vf= \LL_\e\vf\ ,
\eqn{susy-alg}
for all fields $\vf$,
where $\LL_\e$ is the Lie derivative along the vector $\e$.

In order to check the possible invariance of the action
under the vector supersymmetry,
we first note that this is trivially the case for the gauge invariant 
part \equ{inv-action}
of the total action  \equ{tot-action}. 
Next, since the gauge fixing 
part \equ{gf-action}  is a BRST variation:
\eq
S_{\rm gf}= -\dintm d^4x\sqg g^{\m\n} 
\dsum{i=1,2}{} s\lp\pa_\m\BC_i \AA^i_\n\rp \ ,
\eqn{gf-ac}
we can use the anticommutation relation \equ{susy-alg} 
and thus write
\eq
\dse S = -\dintm d^4x\sqg g^{\m\n} 
\dsum{i=1,2}{} \LL_\e \lp\pa_\m\BC_i \AA^i_\n\rp \ .
\eqn{d-susy-ac}
Partial integrations\footnote{We assume allthrough the absence 
of boundary terms contributions.} then yield
\eq
\dse S = \dintm d^4x\sqg 
  \lp \LL_\e g^{\m\n} + \nablag_\la\e^\la g^{\m\n} \rp
\dsum{i=1,2}{} \lp\pa_\m\BC_i \AA^i_\n\rp \ ,
\eqn{d-susy-action}
where $\nablag_\m$ is the covariant derivative with respect to 
the background metric $g_{\m\n}$.
The first conclusion is that, generically, the vector
supersymmetry transformations \equ{vec-susy} are not 
an invariance of the theory. However they will indeed represent
an invariance if and
only if the parenthesis in the integrant of 
the right-hand side of \equ{d-susy-action}
vanishes:
\eq
\LL_\e g^{\m\n} + \nablag_\la\e^\la g^{\m\n} = 0\ ,
\eqn{susy-cond}
which is easily shown to be equivalent to the condition 
that the vector $\e$ be a Killing vector field of the 
background metric:
$g^{\m\n}$:
\eq
\LL_\e g^{\m\n}= 0\ ,\quad\mbox{or:}\quad
   \nablag_\m\e_\n + \nablag_\n\e_\m = 0\ .
\eqn{killing}
In such a case, the vector supersymmetry invariance may be 
expressed by the functional identity (still using the notation
\equ{notation})
\eq
\dse S \equiv \dintm d^4x\, \e^\m \lp \dfud{}{\xi^\m} 
+ \dsum{i=1,2}{} \pa_\m \BC_i\dfud{}{B_i} \rp S = 0\ .
\eqn{susy-w-i}
It is illustrative to consider a flat background metric, e.g.
the Minkowski one: $g_{\m\n}$ $=$ $\eta_{\m\n}$. In this case  
the general solution 
of the condition \equ{killing} reads
\eq
\e^\m = a^\m + b^{\m\n}x_\n\ ,\quad \mbox{with}\quad
a^\m\,,\  b^{\m\n}=-b^{\n\m}\quad \mbox{constants}\ .
\eqn{susy-poinc}
The right hand side of the anticommutator in \equ{susy-alg} is then an
infinitesimal rigid Poincar\'e transformation of parameters 
$a^\m$ and $b^{\m\n}$.

Note that one could have derived the identity \equ{susy-w-i}
directly from the equation \equ{xi-eq}
for the  diffeomorphism ghost $\xi$, integrated with 
the vector field $\e$, and performing some partial 
integations. Thus we clearly see how the vector supersymmetry
is linked to the diffeomorphism ghost equation. 
In this respect the situation differs from the one
encountered in Yang-Mills topological theories (Chern-Simons, BF), 
where there is no such narrow
relation between a ghost equation and the vector supersymmetry
-- although they both hold in a Landau type gauge, too.

Another difference with the topological Yang-Mills case is that the weaker
condition \equ{killing} for supersymmetry invariance
 holds in the present case, whereas it reads  there
$\nablag_\m\e_\n=0$: the superalgebra \equ{susy-alg} closes there on 
isometries of the translation type only - true translations instead of
general Poincar\'e transformations for a flat
background metric.

It is known that in the Yang-Mills case the vector symmetry 
operator may be expressed in the form of 
the so-called operator $\d$ of 
Sorella~\cite{sorella-delta} used to construct the invariants 
of the theory, and characterized by the 
algebraic relation
\eq
[\d,s] = d\ ,
\eqn{delta-s-com}
where $d$ is the exterior derivative. In the present case, too, 
there exists~\cite{brs-grav-wien1,brs-grav-wien2} 
such an operator $\d$. And,
remarkably, it is linked to our vector supersymmetry, 
hence to the diffeomorphism ghost equation, in the 
following way. Considering the supersymmetry
transformation rules \equ{vec-susy} for a
constant\footnote{Vector supersymmetry invariance will then hold
for a flat Minkovskian background metric.}
  vector field $\e$, we define the action 
of the operator $\d$ as given by these transformations, with
$\e^\m$ replaced by the differential $dx^\m$:
\eq\ba{l}
\delta\xi^\m = dx^\m\ ,\es
\delta B_i = d\BC_i\ , \es
\delta\vf = 0\ , \quad \vf \not= \xi^\m\,,\ B^i  \ ,
\ea\eqn{delta}
which is the result of~\cite{brs-grav-wien1,brs-grav-wien2}
 -- up to the action on the Lagrange multiplier fields 
$B_i$, not considered there. We can easily check 
the commutation rule \equ{delta-s-com}, and also that $\d$ 
commutes with $d$. Note that, as in~\cite{brs-grav-wien2},
we can write the first 
of eqs.~\equ{delta} in a coordinate independent way as
\eq
\d \eta^I = E^I\ ,
\eqn{delta-eta}
where $\eta^I\equiv E^I_\m\xi^\m$ is the ``tangent space 
translation ghost''~\cite{BRS-grav,brs-grav-wien2}.

Before concluding, we could 
ask for the role of the Lorentz ghost equation, 
the second of eqs.~\equ{ghosteqs} The answer is much the 
same as in ordinary gauge theories for the Yang-Mills 
ghost equation~\cite{ghost-eq}. 
Integrating the Lorentz ghost equation 
in space-time, integrating by part and using the 
gauge conditions \equ{gauge-cond}, we obtain
\eq
\d_{IJ}^{({\rm SL})}S \equiv \dintm d^4x \lp \dfud{}{\om^{IJ}} 
 - \bxi_I\dfud{}{\la^J} + \bxi_J\dfud{}{\la^I}
 - \bom_{IK}\dfud{}{b^J{}_K} + \bom_{JK}\dfud{}{b^I{}_K} \rp S
  = 0 \ ,
\eqn{eq-ghost-om}
which is very similar to the result of~\cite{ghost-eq}, 
the Yang-Mills 
gauge invariance being now replaced by the local Lorentz 
invariance. One also may check the anticommutation rule
\eq
\{ s, \d_{IJ}^{({\rm SL})} \} = \d_{IJ}^{({\rm L})}\ ,
\eqn{ghost-lorentz-alg}
where the right-hand side is an infinitesimal generator of rigid
Lorentz transformation.

\section{Conclusions}

We have found a direct relation between the diffeomorphism 
ghost equation (the first of eqs. \equ{ghosteqs}) and the 
existence of a vector supersymmetry \equ{vec-susy}
 -- or of the Sorella operator $\d$ \equ{delta}, \equ{delta-eta}. 
This appears to be a characteristic features of
theories invariant under 
``active diffeomorphisms'', i.e. diffeomorphisms 
which act on the dynamical fields only\footnote{In a quantum 
context, these are diffeomorphisms acting quantum mechanically
on the field operators -- vierbein, connection and matter 
fields~\cite{gaul-rovelli}.}. In such theories, the 
diffeomorphism ghost $\xi$ is a dynamical field,
which actually means that it obeys an equation of motion.

Our results do not depend on the specificity of the invariant action
taken to define the theory. They obviously also hold in the case of a 
self-dual connection $A^{IJ}$, in which case the 
action \equ{inv-action} is that of 
Ashtekar~\cite{ashtekar,peldan,baez-knots}. As we have mentionned, the
presence of minimally coupled matter is allowed, as well as other type
of actions, provided they share the same ``topological-like'' character,
i.e. provided they are constructed with the vierbein and connection as
independent variables, and with invariance under active diffeomorphisms,

We have also emphasized the differences of the present
diffeomorphism invariant theory with respect
to the topological theories for Yang-
Mills fields, such as the role of the ghost equation, but also the
existence of more supersymmetry thanks to a weaker condition of 
invariance.

During completion of this work, the author became aware of a 
recent preprint~\cite{ggps} -- published by now --
which gives an alternative derivation of vector supersymmetry in 
topological theories. This method may be applied to the gravitational
case, too -- and has been indeed applied in the published version 
of~\cite{ggps}.

\subsubsection*{Acknowledgments}
The author is very grateful to Fran\c cois Gieres for critical remarks 
and for
having communicated to him results prior to publication. 
Thanks are due to one of the referees for suggestions of 
references and useful comments.



\begin{thebibliography}{99}
\bibitem{peldan} P. Peld\'an, ``Actions for Gravity, with Generalizations: 
A Review'',\\ \cqg{11}{1994}{1087} and e-print gr-qc/9305011;
\bibitem{baez-knots} J. Baez, ``Strings, Loops, Knots and Gauge Fields'',
in {\em ``Knots and Quantum Gravity''}, p.133 (edited by John Baez, 
Clarendon Press, Oxford, 1994);
\bibitem{baez-foam} J.C. Baez, 
``An Introduction to Spin Foam Models of
$BF$ Theory and Quantum Gravity'',
e-print gr-qc/9905087;
\bibitem{witten} E. Witten, \cmp{117}{1988}{353},\\ 
\cmp{118}{1988}{411};
\bibitem{gaul-rovelli} M. Gaul and C. Rovelli, ``Loop Quantum Gravity
and the Meaning of Diffeomorphism Invariance'', Lectures presented at
the {\em 35th Karpacz Winter School on Theoretical Physics: From
Cosmology to Quantum Gravity}, 1999, e-print gr-qc/9910079;
\bibitem{rovelli-living} C. Rovelli, ``Loop Quantum Gravity'', 
{\em Living Reviews on Relativity} 1998-1,
http://www.livingreviews.org/ and e-print gr-qc/9710008;
\bibitem{brst} C. Becchi, A. Rouet and R. Stora,
  {\em Phys. Lett.} B52(1974)344, \cmp{42}{75}{127}\\
I.V. Tyutin, ``Gauge invariance in field theory and
         statistical mechanics``,
       {\em Lebedev preprint FIAN, $n^0$ 39 (1975), unpublished};
\bibitem{alg-ren}  O. Piguet and S.P. Sorella, {\em ``Algebraic
Renormalization''}, Lecture Notes in Physics, m28, Springer-Verlag 
(Berlin-Heidelberg), 1995;
\bibitem{cs-lucch-pig} C. Lucchesi and O. Piguet,
 \np{B381}{1992}{281};
\bibitem{vec-susy} F. Delduc, F. Gieres and S. P. Sorella,
                   \pl{B225}{1989}{367};
\bibitem{UV-finiteness-vector-SUSY}  F. Delduc, C. Lucchesi, 
O. Piguet and S. P. Sorella,\\ \np{B346}{1990}{313};
\bibitem{ghost-eq} A. Blasi, O. Piguet and S. P. Sorella,
            \np{B356}{1991}{154};
\bibitem{ghost} R. Feynman, {\em Acta Physica Polonica}
24\,(1963)\,697;\\
B. DeWitt, \pr{162}{1971}{1195,1239};\\
L.D. Faddeev and V.N. Popov, \pl{B25}{1967}{29};
\bibitem{sorella-delta}  S.P. Sorella, \cmp{157}{1993}{231};
\bibitem{brs-grav-wien1} M. Werneck de Oliveira and S.P. Sorella.
\ijmp{A9}{1994}{2979} and e-Print hep-th/9304007;
\bibitem{brs-grav-wien2} 
O. Moritsch, M. Schweda and S. P. Sorella,\\
\cqg{11}{1994}{1225} and e-print hep-th/9310179;\\
P.A. Blaga, O. Moritsch, M. Schweda, 
T. Sommer, L. T\u ataru  and H. Zerrouki, 
\pr{D51}{1995}{2792} and e-print hep-th/9409046;
\bibitem{emery} S. Emery, O. Moritsch, M. Schweda, T. Sommer, 
H. Zerrouki,
\hpa{68}{1995}{167}, e-print hep-th/9503192;
\bibitem{ashtekar} A. Ashtekar, ``Lectures on Non-perturbative 
Canonical Quantum Gravity'', World Scientific (Singapore) 1991;
\bibitem{BRS-grav} J. Thierry-Mieg, \jmp{21}{1980}{2834},\\
\nc{56A}{1980}{396}, \pl{B147}{1984}{430}; \\ 
F. Langouche, T. Sch\"ucker and R. Stora,
\pl{B145}{1984}{342};
\bibitem{dewitt} B. DeWitt, ``Space-time Approach to Quantum Field
Theory'', in {\em Relativity,
   Groups and Topology II, Les Houches, 1983, ed. B.S. DeWitt and R. Stora
  (North Holland, Amsterdam, 1984)}, pp. 381--738. 

\bibitem{ggps} F. Gieres, J. Grimstrup, T. Pisar and M. Schweda,
``Vector supersymmery in topological theories'',
J. High Energy Phys. (JHEP) 06 (2000) 018; former version 
available as e-print hep-th/0002167.

\end{thebibliography}
\end{document}